\documentclass[epjST]{svjour}
\usepackage{graphics}
\begin{document}
\title{Control of dynamical instability in semiconductor quantum nanostructures diode lasers: role of phase-amplitude coupling}
\author{Pramod Kumar\inst{1}\fnmsep\thanks{\email{pramod@iisermohali.ac.in, pk6965@gmail.com}} \and Fr\'{e}d\'{e}ric Grillot\inst{2}}
\institute{Femtosecond Laser Laboratory, Department of Physical Sciences, Indian Institute of Science Education and Research (IISER) Mohali-140306,
 Punjab, India. \and T\'{e}l\'{e}com ParisTech, Ecole Nationale des
 Sup\'{e}rieure des T\'{e}l\'{e}communications de Paris, Laboratoire CNRS 5141, 46 rue Barrault, 75634 Paris Cedex 13, France}
\abstract{
We numerically investigate the complex nonlinear dynamics for two independently coupled laser systems consisting of (i) mutually delay-coupled edge emitting diode lasers and 
(ii) injection-locked quantum nanostructures lasers. A comparative study in dependence on the dynamical role of $\alpha$ parameter, which determine the
phase-amplitude coupling of the optical field, in both the cases is probed. The variation of $\alpha$ lead to conspicuous 
changes in the dynamics of both the systems, which are characterized and investigated as a function of optical injection 
strength $\eta$ for the fixed coupled-cavity delay time $\tau$.  Our analysis is based on
 the observation that the cross-correlation and bifurcation measures unveil the signature of enhancement of 
amplitude-death islands in which the coupled lasers mutually stay in stable phase-locked states. In addition, we provide a qualitative understanding of
 the physical mechanisms underlying the observed dynamical behavior and its dependence on $\alpha$. The amplitude death and the existence 
of multiple amplitude death islands could be implemented for applications including diode lasers stabilization.
} %end of abstract
\maketitle
\section{Introduction}
\label{intro}
The semiconductor diode lasers are known to be very sensitive to the external optical perturbations 
such as, optical self-feedback, optoelectronic feedback, optical injection. When the diode
 laser is subjected to optical injection by another diode laser then the radiation emitted from mutually 
delay-coupled diode lasers is well behaving, well understandable and well classifiable in terms of complex 
nonlinear dynamics \cite{kumar08}. On the one hand these dynamical instabilities are undesired features and disturb 
the many applications where one needs the constant stable power but one the other hand they may allow 
 new methods for the secure communications using chaos synchronization. So the systematic study and control
 of these nonlinear dynamics provide fundamental insights into the underlying physics of the system \cite{resmi11}. On the 
basis of which one can redesign the device\cite{pal11} or improve the processing, or simply 
exploit the dynamical performance of a system to one’s advantage.
Amplitude death \cite{kumar09} is one of the collective fascinating phenomena \cite{pune11}, where coupled lasers drive each 
other to a fixed point and stop the oscillatory dynamics. The coupling induced stabilization is known 
to occur either by changing the stability of unstable fixed points which already exists in the 
absence of coupling or the stationary state which can be created newly by the coupling. Because of very diverse time scales involved in the dynamics, the
 routes by which the two coupled lasers reach their stable states in the presence of finite-delayed interaction
 remain poorly understood . In this sprit, we have addressed the question whether these dynamical behaviors of a coupled diode lasers and particularly  quantum nanostructures 
based diode lasers can be controlled via amplitude death and the kind of dynamics one encountere during the route to reach the ultimate stable state.  
\section{The Model for the delay-coupled diode lasers system}
For numerical simulations, the time evolutions of the complex electric field $E_{1,2}(t)$ and the  effective carrier 
density $N_{1,2}(t)$ ($n_{th}-n_{1,2}$) in the laser medium for each delay-coupled diode lasers are modeled.
\begin{eqnarray}
\frac{dE_{1}}{dt}&=&(1+i\alpha)N_{1}(t)E_{1}(t)+\eta ~\mathrm{e}^{-i\omega_{2}\tau}E_{2}(t-\tau),\label{eq1}\\
T\frac{dN_{1}}{dt}&=&J_{1}-N_{1}-\left[2N_{1}+1\right]\left| E_{1}(t)\right|^{2},\label{eq2}\\
\frac{dE_{2}}{dt}&=&(1+i\alpha)N_{2}(t)E_{2}(t)+\eta ~\mathrm{e}^{-i\omega_{1}\tau}E_{1}(t-\tau),\label{eq3}\\
T\frac{dN_{2}}{dt}&=&J_{2}-N_{2}-\left[2N_{2}+1\right]\left| E_{2}(t)\right|^{2},\label{eq4}
\end{eqnarray}
In order to analyze the complex dynamical behaviour of two mutually delay-coupled diode lasers system (as shown in figure 1), the LK 
equations can be written in a standard normalized form \cite{kumar08}:
where $\eta$ is the effective coupling strength, i.e., the fraction of light of one laser injected into the other laser
 and vice versa, $J_{1,2}$'s are the effective current densities (with the threshold value subtracted out), $T$ is 
the ratio of the carrier lifetime to the photon lifetime, $\alpha$ \cite{henry82} is the linewidth enhancement factor or phase-amplitude coupling factor, and $\tau$ is
 the coupled cavity time taken by the light to cover the distance between the two lasers system. $\omega_{1,2}$ are the optical angular frequencies of the 
free running diode lasers $1$ and $2$. $E_{1,2}(t-\tau)$ are the fields delayed by one coupling time $\tau=L/c$ and $\omega_{1,2}\tau$ are the coupling
phase detuning. For simplicity, we have taken two similar diode lasers for which the frequency detuning between them is assumed to be very small or nearly zero. 
In order to mimick the real experimental situation, we have introduced some small noise in our model. We have done numerical integration of the above 
equations by using Runge-Kutta fourth-order method with a step size $= \tau /n$, where $n=1000$ is chosen 
based on the response time of photodiode in experimental setup\cite{kumar08}. The dimensionless parameters are taken as $J_{1,2}=0.165, 175$ and $T=1000$ \cite{kumar08}.
In order to scan and understand the different dynamical regimes, we use equation $5$ for correlation measurement \cite{kumar09}.We observe 
the route to amplitude death of low frequency complex dynamics in
 the output power of a diode laser (slave) when subjected to optical injection from another diode laser (master).
In the context of a system of two delay-coupled diode lasers and Quantum dot nanostructures lasers, we emphasize the effect 
of $\alpha$ on the control of complex dynamical instabilities near the phase flip transition regimes as a function of coupled-cavity
 time delay $\tau$ and the optical injection strength $\eta$. Shifting of different phase-correlated dynamics, such as phase-flip 
bifurcation \cite{kumar08} and strange bifurcation \cite{kumar09} was observed in our previous work when $\eta$ is varied for a particular $\tau$. In
 this present work, we show that these phase flip transitions does not occur abruptly at a particular 
value of $\eta$ and $\alpha$.
\label{sec:1}
\begin{figure}
\begin{center}
\resizebox{0.65\columnwidth}{!}{%
\includegraphics{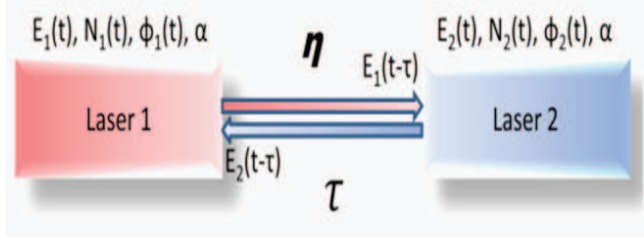} }
\caption{(Color online) Schematic diagram of a Delay-coupled
diode laser system.}
\label{pfig:1}
\end{center}       % Give a unique label
\end{figure} 
 Instead we find a coupling strength region around the phase flip transition where the 
co-existence of multi-attractors occurs as shown in figure $2$ and figure $3$. We show that the phase 
flip bifurcation occurs from in-phase amplitude death to anti-phase amplitude death and  the strange bifurcation 
occur from anti-phase to in-phase transition regimes \cite{kumar12}.
 The existence of multiple attractors near the regime of 
strange bifurcation (as shown in fig. 2. and fig. 4.) has raised the issue of whether the $\alpha$ plays a crucial role or not. We also study the effect 
of $\eta$ along with the phase-amplitude coupling factor on the dynamics in the amplitude death regime and extends the study for
 nanostructures quantum dot lasers. One of the Key features of semi-conducteur lasers is their $3-dB$ modulation 
bandwidth, which is limited due to the presence of strong relaxation oscillation damping rates. This work also
 theoretically focuses on the impact of strong injection in a quantum nanostructure semiconductor laser with 
large variation of the phase-amplitude coupling factor through the non-linear dynamics and the modulation response
 at zero-detuning.The combination of the strong injection, optimized phase-amplitude coupling factor and the zero-detuning case shows the
 possibility to reach the stable state, which is fully suitable for laser control  applications.
\section{Influence of $\alpha$ in delay-coupled edge-emitting diode lasers and injection-locked quantum nanostructures lasers}
\label{sec:2}
The linewidth enhancement factor $\alpha$ describes the coupling between the real and imaginary parts of the material 
refrective index of the gain medium, and is given by the ratio of their derivatives with respect to the carrier density. Its value 
is known to depend on the carrier concentration, photon wavelength, inhection current and  temperature. 
For quantum nanostructures based diode lasers (quantum dot lasers ), the value of $\alpha$ can in
 fact be controlled by the desirable design of the device structure or by the filtered optical feedback or injection techniques. Unlike the quantum dot lasers, the $\alpha$ in bulk semiconductor lasers can be modified just by
varying the specific combination of control 
parameters in a particular range.  In order to analyze the role of $\alpha$ in the various dynamical regimes 
of mutually delay-coupled diode lasers, we use a normalized cross-correlation function defined as 
\begin{equation}
C=\frac{\langle(P_{1}(t)-\langle P_{1}(t)\rangle)(P_{2}(t)-\langle P_{2}(t)\rangle)}{\sqrt{\langle (P_{1}(t)-\langle P_{1}(t)\rangle)^{2}\rangle \langle (P_{2}(t)-\langle P_{2}(t)\rangle)^{2}\rangle}}.\label{eq14}
\end{equation}
Note that the measured output powers of the lasers, $P_{1, 2} \equiv |E_{1, 2}|^2$ do not explicitly depend on 
the optical phases $\phi_{1, 2}$ of the electric fields. We wish to probe the role of phases of optical fields through $\alpha$ in 
the intensity cross-correlation near the phase bifurcation or transitions of two mutually delay-coupled diode lasers.
\begin{figure} 
\begin{center}
\resizebox{0.65\columnwidth}{!}{%
\includegraphics{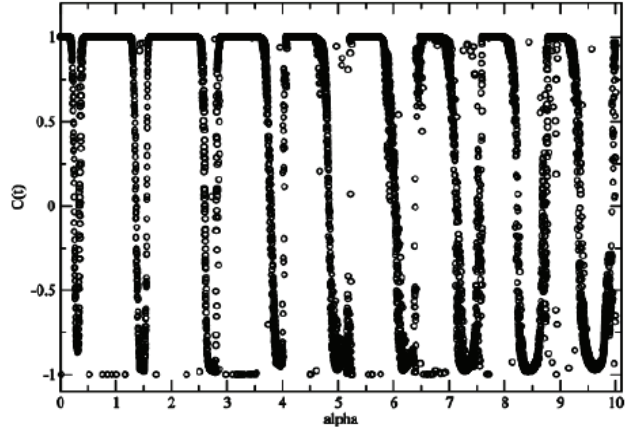} }
\caption{Plot of the cross-correlation $C(t)$ versus $\alpha$ for a fixed time
delay $\tau=14$ (in units of cavity photon lifetime).}
\label{pfig:2}       % Give a unique label
\end{center}
\end{figure}
We start by analyzing the modulation properties of the coupled laser system from the rate equations $(1-4)$. In Figure $4$, the
symbol AD represents the signature of ultimate death state while BS and MS represent the Bistability and multistability between in-phase amplitude-death island and
anti-phase amplitude-death island respectively. Recall that the phase-space plot of this ultimate death state is represented as AD in figure $5$, and the MD dynamics is
shown in figure $6$.
\begin{figure}
\begin{center}
\resizebox{0.65\columnwidth}{!}{%
\includegraphics{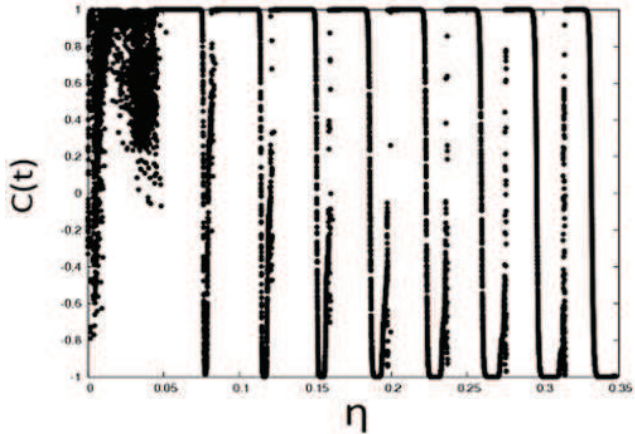} }
\caption{Plot of the cross-correlation $C(t)$ versus coupling strength,
$\eta$ for a fixed time delay $\tau=14$ (in units of cavity photon lifetime).}
\label{pfig:3}       % Give a unique label
\end{center}
\end{figure}
\begin{figure}
\begin{center}
\resizebox{0.75\columnwidth}{!}{%
\includegraphics{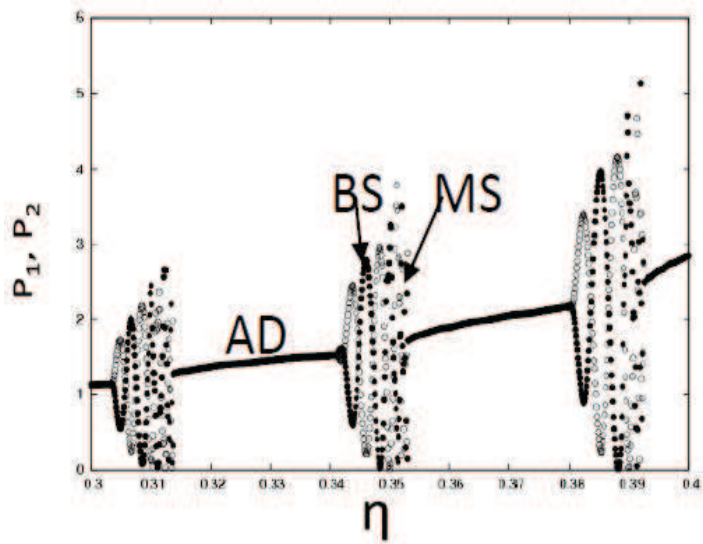} }
\caption{Bifurcation Plots of laser output powers $P_{1}$ (open circles)
and $P_{2}$ (filled circles) versus coupling strength $\eta$ for a fixed time
delay $\tau=14$ (in units of cavity photon lifetime).}
\label{pfig:4}       % Give a unique label
\end{center}
\end{figure}
From the cross-correlation measures, we have observed a unusual phase transitions from anti-phase to inphase 
state  (as shown in fig. 3) via strange bifurcation as the control parameter, $\alpha$ is varied for a particular value of $\eta$. So the phase-amplitude coupling 
factor induced multistability has been observed in the system as shown in fig. 4. The amplitude death attractors is shown in fig. 5 within the AD regimes of fig. 4.
The three different attractors (MS regimes in fig. 4.), depending on the initial conditions, are shown in fig. 6. as the $\eta$ scaned back and forth.  
\begin{figure}
\begin{center}
\resizebox{0.75\columnwidth}{!}{%
\includegraphics{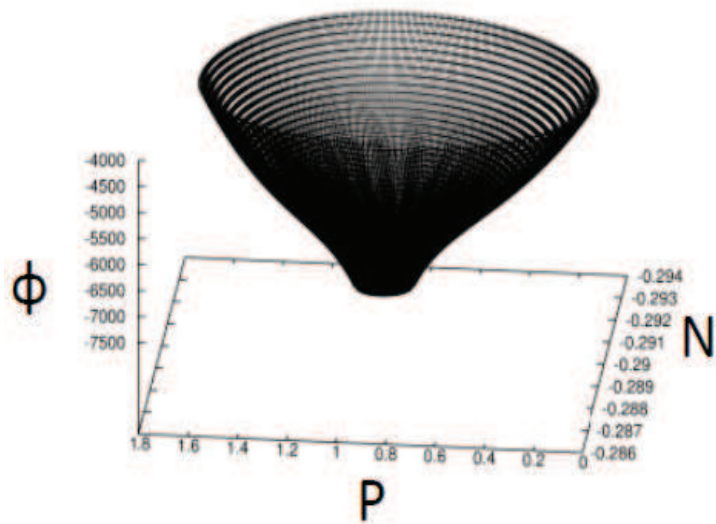} }
\caption{Plot of Amplitude death attractor of the AD state shown
in figure $4$ in the parameter space of $P(t)$, $N(t)$, $\phi(t)$ for a fixed
time delay $\tau=14$.}
\label{pfig:5}       % Give a unique label
\end{center}
\end{figure}
\begin{figure}
\begin{center}
\resizebox{0.65\columnwidth}{!}{%
\includegraphics{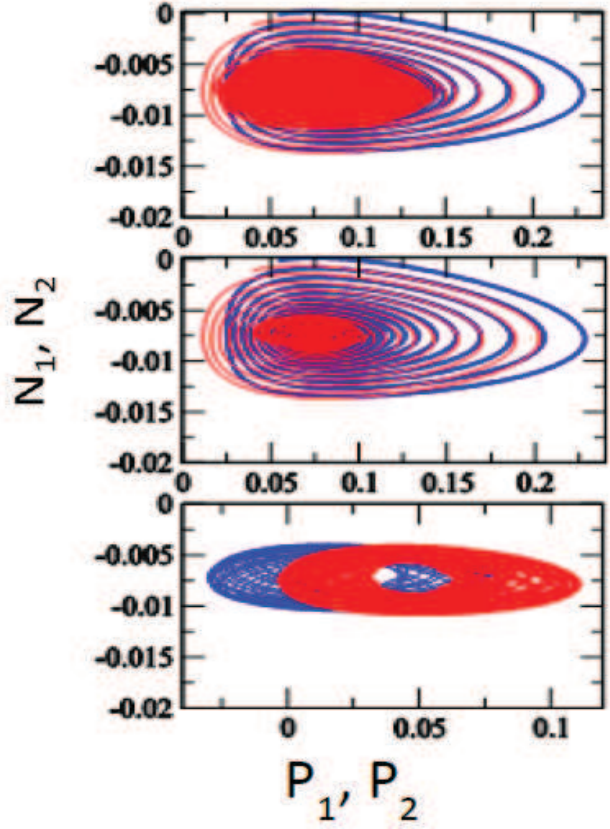} }
\caption{(Color online) phase space plots of the multi-attractors
dynamics (MS state in figure 4.) between two stable locking states
as the initial condition are varied for fixed $\eta$, $\tau$ and initial
transients of $10^{5}$ data points are discarded.}
\label{pfig:6}       % Give a unique label
\end{center}
\end{figure}
Injection-locking of semiconductor lasers is one of the most attractive research topics since 
this method induces superior improvement in the high-speed characteristics of directly modulated 
lasers such as increasing the modulation bandwidth, suppressing nonlinear distortion, relative 
intensity noise \cite{sim95}, mode hopping and reducing chirp \cite{nad09}. Previous work has focused on realizing 
high modulation bandwidths and associated design strategies, analyzed the modulation properties of 
the coupled system in the spectral domain, and numerically investigated the modulation response of
 the injection-locked system \cite{grillot08}. In order to understand the limiting factors in an injection-locked system 
it is important to investigate the governing theory that can be obtained by properly modeling the impact 
of characteristic parameters such as the so-called linewidth enhancement factor ( $\alpha$-parameter). A 
dimensionless volume averaged normalized approach to theoretically evaluate the nonlinear dynamics as a 
function of the injected field ratio and/or the detuning 
frequency for varied slave laser bias cases can be described as follows \cite{ern96}: 
\begin{eqnarray}
\frac{dY}{d\tau}&=&ZY+\epsilon Y(Y^{2}-P)+\eta\cos\theta,\label{eq5}\\
\frac{d\theta}{d\tau}&=&\alpha Z-\alpha\epsilon (Y^{2}-P)-\frac{\eta}{Y}\sin\theta-\Delta\Omega,\label{eq6}\\
T\frac{dZ}{d\tau}&=&P-Z-Y^{2}(1+2Z-2\epsilon Y^{2}+2\epsilon P)+\eta\cos\theta,\label{eq7}
\end{eqnarray}
with $Y$, describing 
the normalized field magnitude and $Z$ the normalized carrier density. The $T$-parameter 
is the ratio of the cavity decay rate to the spontaneous carrier relaxation rate. Parameter $P$ is 
proportional to the pumping current above threshold while coefficient $\epsilon$ accounts for the nonlinear 
carrier contribution to relaxation rate. The detuning and phase offset between the master and the 
slave are denoted by $\Delta\Omega$ and $\theta$, respectively. The normalized injection strength is $\eta=\eta_{0}(P/\gamma_{c})^{1/2}$with $\gamma_{c}$ the cavity 
decay rate and $\eta_{0}$ the maximum injection strength. In solving the coupled normalized differential 
equations, the normalized field magnitude $Y$ is not at steady state, and is thus represented as a 
dependent term in the normalized field magnitude and phase rate equations. In what follows, this 
model is used for “the stability analysis of a quantum dot laser.
\begin{figure}
\begin{center}
\resizebox{0.75\columnwidth}{!}{%
\includegraphics{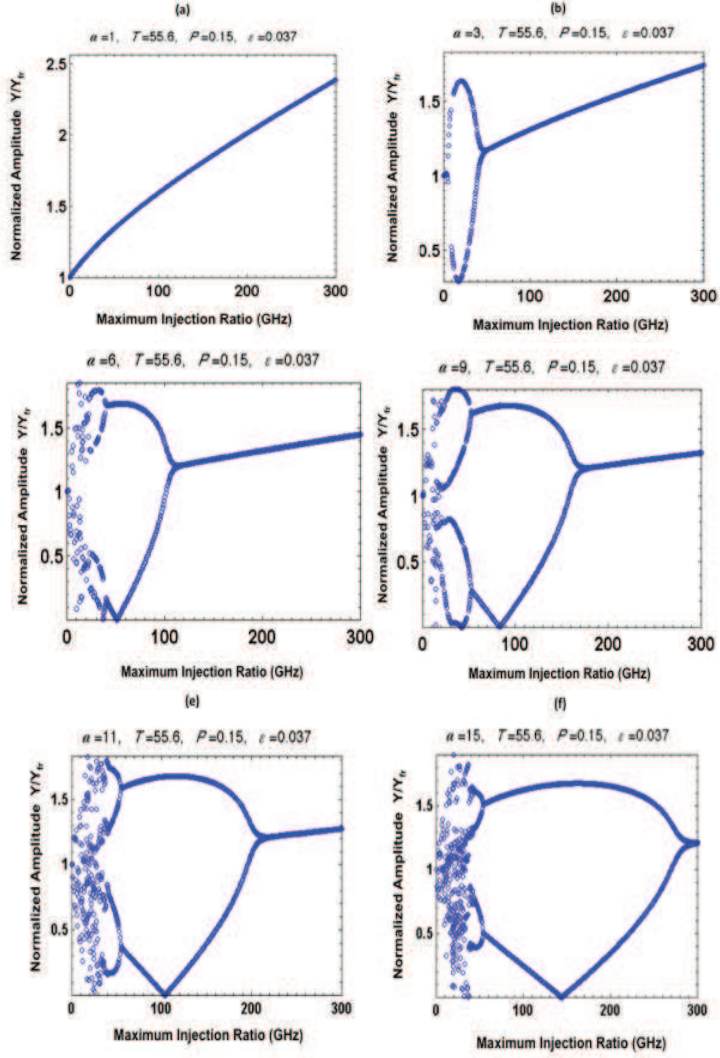} }
\caption{Calculated bifurcations diagrams of a quantum
dot laser with a variable linewidth enhancement factor.}
\label{pfig:7}       % Give a unique label
\end{center}
\end{figure} 
The quantum dot laser under study is a ridge waveguide with $500$-µm cleaved cavity length. The 
ground state (GS) emission wavelength is at $1560 nm$ \cite{nad09}.Because of the carrier filling in the 
lasing and non-lasing higher energy levels, gain compression effects are much larger as compared to
 bulk or quantum well materials\cite{grillot08}. Consequently, it has been shown that the $\alpha$-factor in quantum dot 
lasers cannot be considered as a constant parameter since it strongly varies from one laser to another 
and also with optical output power. Figure $7$ shows the bifurcation diagrams calculated at a constant 
pump current but for different values of the $\alpha$-parameter ranging from 1 to 15(assuming$\Delta\Omega=0$). The objective 
of the calculations is to show the effects of a large $\alpha$-factor on the laser’s stability. Numerical results 
point out that taking into account such variations reveals strong modifications in the bifurcation 
diagram. On one hand, at low $\alpha$ (case (a)), the laser is always stable while for $\alpha=3$ (case (b)), period one oscillation
 starts occurring. On the other hand, for $\alpha>3$ (cases (c), (d), (e) and (f)),the bifurcation diagram exhibits chaos 
(at low injection ratio) followed by a cascade of periodic regimes converging to a stability 
area (at very high injection ratio). From a general point of view, simulations reveal that the 
smallest the $\alpha$-parameter, the better. As far as the engineering point of view is concerned, these 
results are very important for the optimization of the microwave properties. Although the optical 
injection is used to purify the relaxation frequency as well as the modulation bandwidth, these numerical
 results point out that a large $\alpha$-factor is detrimental for the laser’s stability and can induce severe 
degradations in the microwave properties.As a conclusion, in order to maintain a wide stability area with 
optical injection associated to good microwave properties, a low $\alpha$-factor is mandatory in quantum dot laser.
\section{Conclusion}
\label{sec:4}
This new concept stabilizes the laser emission by directly controlling the nonlinearity and, 
consequently, enhances the stability properties of the system. In general, semiconductor lasers with a 
sufficiently low $\alpha$ would be most interesting for practical applications due to the possibility 
of chirpless operation, and the insensitivity to delayed optical feedback or injection. The different
 dynamics and the strange bifurcation among them is investigated as a function of coupled-cavity time 
delay $\tau$ and the optical injection strength $\eta$. The correlation measure gives the signature of variation in 
amplitude death islands of complex dynamics of the delay-coupled lasers. The shrinkage of ‘in-phase death state’ and 
enlargement of ‘out-of-phase death state’ are observed (as shown in fig. 2 and fig. 3) and analyzed when $\alpha$ and $\eta$ are varied. We provide detailed
 information about the effect of the variation of $\alpha$ on the dynamics of the system over wide ranges of experiment relevant 
parameters. In particular, we give numerical evidence that the stability of the system increases with 
decreasing $\alpha$. This last point is particularly predominant in a quantum dot laser for which a larger $\alpha$-factor is usually observed.

\end{document}